\documentclass[a4paper, 11pt]{article} 
\pdfoutput=1

\usepackage{subfig}
\usepackage{graphicx}

\newcommand{\ntup}[1]{\ensuremath{n}-tuplet#1}
\usepackage{siunitx}
\usepackage{url,microtype}
\usepackage[onehalfspacing]{setspace}
\usepackage{jheppub}
\title{Heterogeneous reconstruction of tracks and primary vertices with the CMS pixel tracker}

\author[1]{A.~Bocci,}
\author[2]{M.~Kortelainen,}
\author[1]{V.~Innocente,}
\author[1]{F.~Pantaleo,}
\author[1]{M.~Rovere}

\affiliation[1]{CERN, European Organization for Nuclear Research, Meyrin, Switzerland}
\affiliation[2]{Fermi National Accelerator Laboratory, Batavia, Illinois, U.S.A.}

\emailAdd{felice.pantaleo@cern.ch}

\keywords{GPU, Particle Track Reconstruction, Vertex Reconstruction, Heterogeneous Computing, Patatrack}

\abstract{
The High-Luminosity upgrade of the LHC will see the accelerator reach an instantaneous luminosity of \SI{7d34}{\per\square\centi\metre\per\second} with an average pileup of \num{200} proton-proton collisions. These conditions will pose an unprecedented challenge to the online and offline reconstruction software developed by the experiments. The computational complexity will exceed by far the expected increase in processing power for conventional CPUs, demanding an alternative approach.

Industry and High-Performance Computing (HPC) centres are successfully using heterogeneous computing platforms to achieve higher throughput and better energy efficiency by matching each job to the most appropriate architecture.

In this paper we will describe the results of a heterogeneous implementation of pixel tracks and vertices reconstruction chain on Graphics Processing Units (GPUs). The framework has been designed and developed to be integrated in the CMS reconstruction software, CMSSW. The speed up achieved by leveraging GPUs allows for more complex algorithms to be executed, obtaining better physics output and a higher throughput.

}

\begin{document}
\maketitle
\flushbottom

\section{Introduction}

The High-Luminosity upgrade of the LHC~\cite{apollinari2017high} will pose unprecedented challenges to the reconstruction software used by the experiments due to the increase both in instantaneous luminosity and readout rate. 
In particular, the CMS experiment at CERN~\cite{collaboration2008cms} has been designed with a two-levels trigger system: the Level 1 Trigger, implemented on custom-designed electronics, and the \emph{High Level Trigger} (HLT), a streamlined version of the CMS offline reconstruction software running on a computer farm. A software trigger system requires a trade-off between the complexity of the algorithms running on the available computing resources, the sustainable output rate, and the selection efficiency. 

When the HL-LHC will be operational, it will reach a luminosity of \SI{7d34}{\per\square\centi\metre\per\second} with an average pileup of \num{200} proton-proton collisions. To fully exploit the higher luminosity, the CMS experiment will increase the full readout rate from \SI{100}{\kilo\Hz} to \SI{750}{\kilo\Hz}~\cite{l1triggerTDR:2714892}. The higher luminosity, pileup and input rate present an exceptional challenge to the HLT, that will require a processing power larger than today by more than an order of magnitude.%

This exceeds by far the expected increase in processing power for conventional CPUs, demanding alternative solutions.

A promising approach to mitigate this problem is represented by \emph{heterogeneous computing}. Heterogeneous computing systems gain performance and energy efficiency not by merely increasing the number of the same-kind processors, but by employing different co-processors specifically designed to handle specific tasks in parallel. Industry and High-Performance Computing centres (HPC) are successfully exploiting heterogeneous computing platforms to achieve higher throughput and better energy efficiency by matching each job to the most appropriate architecture.

In order to investigate the feasibility of a heterogeneous approach in a typical High Energy Physics experiment, the authors developed a novel pixel tracks and vertices reconstruction chain within the official CMS reconstruction software CMSSW~\cite{jones2006cmssw}. The input to this chain is represented by RAW data coming out directly from the detector's front-end electronics, while the output is represented by legacy pixel tracks and vertices that could be transparently re-used by other components of the CMS reconstruction.

The results shown in this article are based on Open Data released by the CMS, while the data formats were derived from the CMS Experiment~\cite{cms2019opendata}.

The development of a heterogeneous reconstruction faces several fundamental challenges:
\begin{itemize}
   \item the adoption of a different programming paradigm;
   \item the experimental reconstruction framework and its scheduling must accommodate for heterogeneous processing;
    \item the heterogeneous algorithms should achieve the same or better physics performance and processing throughput as their CPU siblings;
   \item it must be possible to run and validate on conventional machines, without any dedicated resources.
\end{itemize}   

This article is organized as follows: Section~\ref{sec:framework} will describe the CMS heterogeneous framework, Section~\ref{section:PatatrackSequence} will discuss the algorithms developed in the Patatrack pixel track and vertex reconstruction workflow, Section~\ref{sec:results} will describe the physics results and computational performance and compare them to the CMS pixel track reconstruction used at the HLT for data taking in 2018, while Section~\ref{sec:conclusion} will contain our conclusions.


\section{Software framework}\label{sec:framework}

\subsection{CMSSW}

The backbone of the CMS data processing software, \textit{CMSSW}, is a rather generic framework that processes independent chunks of data~\cite{jones2006cmssw}. These chunks of data are called \textit{events}, and in CMS correspond to one full readout of the detector. Consecutive events with uniform calibration data are grouped into \textit{luminosity blocks}, that are further grouped into longer \textit{runs}. 

The data are processed by \textit{modules} that communicate via a C++-type-safe container called event (or luminosity block or run for the larger units). An \textit{analyzer} can only read data products, while a \textit{producer} can read and write new data products and a \textit{filter} can, in addition, decide whether the processing of a given event should be stopped. Data products become immutable (or more precisely, \texttt{const} in the C++11 sense) after being inserted into the event.

During the Long Shutdown 1 and Run 2 of the LHC, the CMSSW framework gained multi-threading capabilities~\cite{jones2014cmssw,jones2015cmssw,jones2017cmssw} implemented with the Intel Threading Building Blocks (TBB) library. The threading model employs task-level parallelism to process concurrently independent modules within the same or different events, multiple events within the same or different luminosity blocks and intervals of validity of the calibration data. Currently the boundary of runs incur barrier-style synchronization point in the processing.
A recent extension is the concept of \emph{external worker}, a generic mechanism to allow producers in CMSSW to offload asynchronous work outside of the framework scheduler.

More details on the concept of external worker and its interaction with CUDA can be found in~\cite{bocci2020bringing}.

\section{Pixel Track and Vertex Reconstruction Strategy}\label{section:PatatrackSequence}

\begin{figure}[!htbp]
    \centering
    \includegraphics[width=0.2\textwidth]{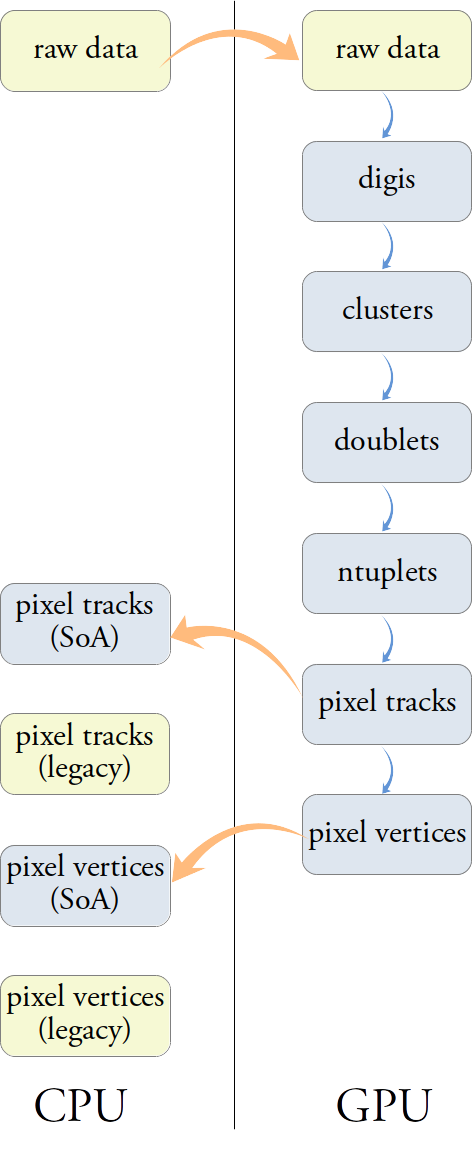}
    \caption{Steps involved in the tracks and vertices reconstruction starting from the pixel ``raw data''.}
    \label{fig:pixeltracksteps}
\end{figure}

Precise track reconstruction becomes more challenging at higher pileup, as the number of vertices and the number of tracks increase, making the pattern recognition and the classification of hits produced by the same charged particle a harder combinatorial problem. To mitigate the complexity of the problem the authors developed parallel algorithms that can perform the track reconstruction on GPUs, starting from the ``raw data'' from the CMS Pixel detector, as will be described later in this section. The steps performed during the tracks and vertices reconstruction are illustrated in Fig.~\ref{fig:pixeltracksteps}. 

The data structures (structure of arrays, SoA) used by the parallel algorithms are optimized for coalesced memory access on the GPU and differ substantially from the ones used by the standard reconstruction in CMS (legacy data formats). The data transfer between CPU and GPU and their transformation between legacy and optimised formats are very time consuming operations. For this reason the authors decided to design a fully contained chain of modules that runs on the GPU starting from the ``raw-data'' and produces the final tracks and vertices as output. While a ``mixed CPU-GPU workflow'' is not supported, for validation purposes the intermediate data products can be transferred from the GPU to the CPU and converted to the corresponding legacy data formats.

\subsection{Local Reconstruction in the Pixel Detector}

\begin{figure}[!tbp]
    \centering
    \includegraphics[width=0.7\textwidth]{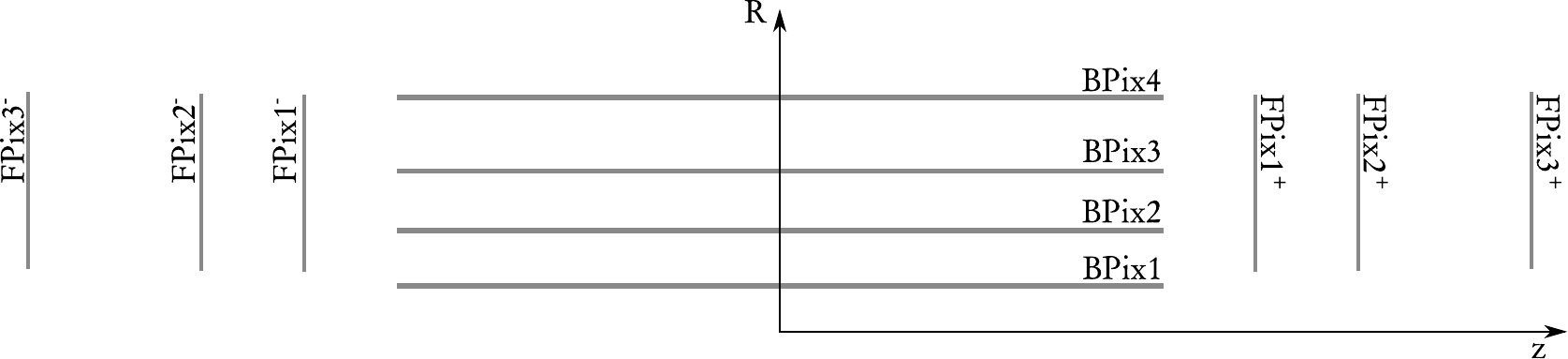}
    \caption{Longitudinal sketch of the pixel detector geometry.}
    \label{fig:pixel_geometry}
\end{figure}

The CMS ``Phase 1'' Pixel detector~\cite{PixelTDR}, installed in 2017, will serve as the vertex detector until the major ``Phase 2'' upgrade for the HL-LHC. It consists of $1856$ sensors of size $1.6$~cm by $6.3$~cm each with $66,560$~pixels, for a total of $124$~million pixels, corresponding to about $2$~$\mathrm{m}^2$ total area. The pixel size is $100$~$\mu m \times 150$~$\mu m$, the thickness of the sensitive volume is $285$~$\mu m$. The sensors are arranged in four ``barrel'' layers and six ``endcap'' disks, three on each side, to provide four-hit pixel coverage up to a pseudorapidity of $|\eta|<2.5$. 
The CMS pixel detector geometry is sketched in Fig.~\ref{fig:pixel_geometry}. The barrel layers extend for $26.7$~cm on each side of the center of the detector. The innermost barrel sensors are located at radius of $2.7$~cm from the beam line, and the farthest at $16.4$~cm.
The forward disks are located between $32$~cm and $48$~cm from the center of the detector along the beam line and cover radii between $4.6$~cm and $16$~cm.
While hermeticity is guaranteed along the azimuthal coordinate thanks to sensor overlaps, gaps exist between sensors along the direction of the beam in the barrel and in the radial direction in the endcaps.
A larger gap exits between the barrel and each endcap.

The analog signals generated by charged particles traversing the pixel detectors are digitized by the read-out electronics and packed to minimize the data rate.
The first step of the track reconstruction is thus the \emph{local reconstruction}, that reconstructs the information about the individual hits in the detector.

During this phase, the digitized information is unpacked and interpreted to create \emph{digis}: each \emph{digi} represents a single pixel with a charge above the signal-over-noise threshold, and contains information about the collected charge and the local row and column position in the module grid. 
This process is parallelized on two levels: information coming from different modules is processed in parallel by independent blocks of threads, while each digi within a module is assigned a unique index and is processed by a different thread.

Neighboring digis are grouped together to form \emph{clusters} using an iterative process. Digis within each module are laid out on a two dimensional grid using their row, column and unique index information. Each digi is then assigned to a thread. If two or more adjacent digis are found, the one with the smaller index becomes the \emph{seed} for the others. This procedure is repeated until all the digis have been assigned to a seed and no other changes are possible. The outcome of the clusterization is a cluster index for each digi: a thread is allocated to each seed; a global atomic counter is increased by all threads, returning the unique cluster index for each seed, and thus for each digi.

Finally, the shape of the clusters and the charge of the digis are used to determine the \emph{hit position} and its uncertainty in the coordinates local to the module as described in  \cite[\S3.1]{Chatrchyan:2014fea}.

\subsection{Building \ntup{s}}

Clusters are linked together to form \ntup{s} that are later fitted to extract the final track parameters.
The \ntup{s} production proceeds through the following steps:
\begin{itemize}
    \setlength\itemsep{0em}
    \item creation of doublets
    \item connection of doublets
    \item identification of root doublets
    \item depth-first-search (DFS) from each root doublet
\end{itemize}

\begin{figure}[!bp]
    \centering
    \includegraphics[width=0.4\textwidth]{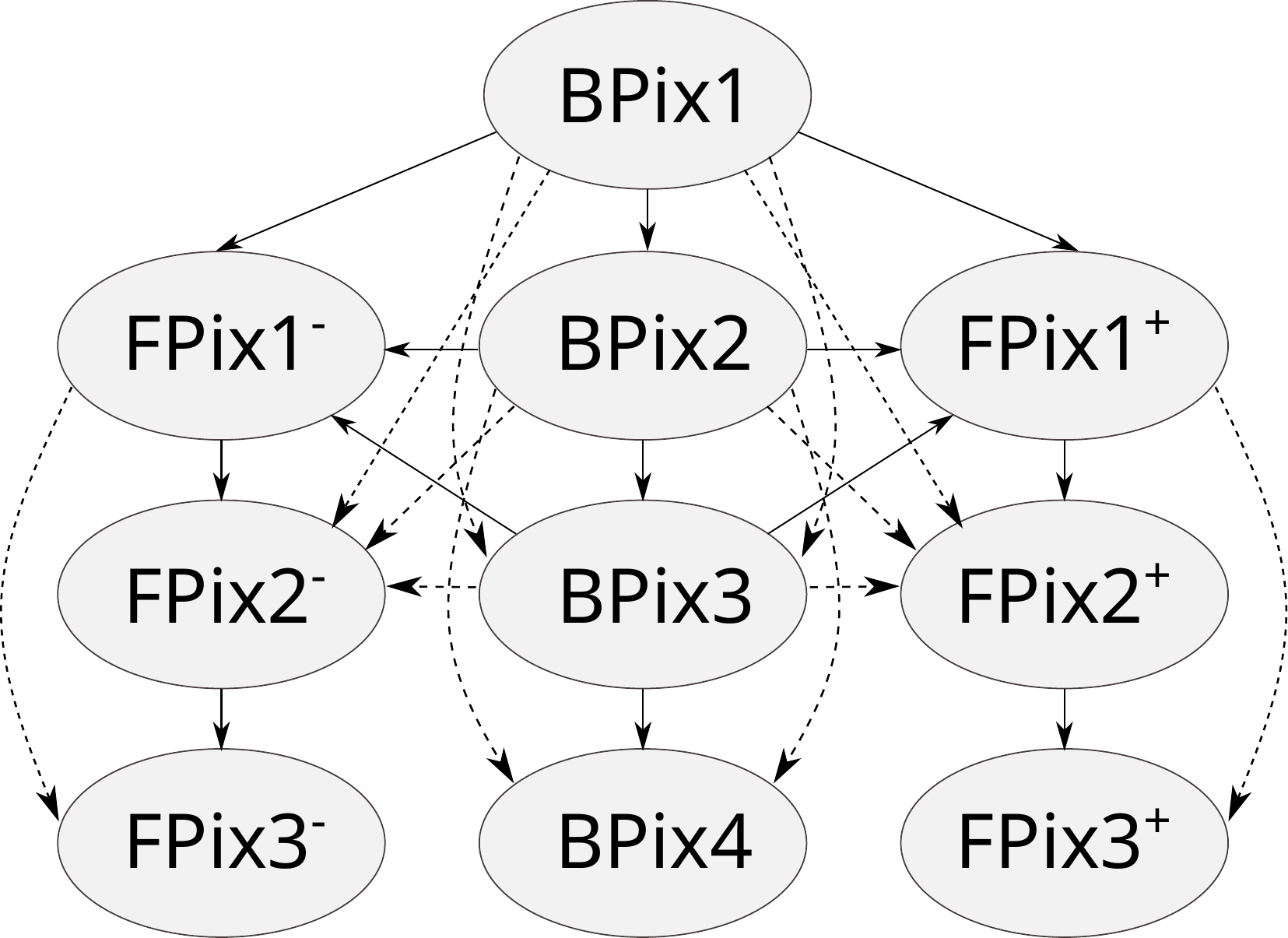}
    \caption{Combinations of pixel layers that can create doublets directly (solid arrow), or by skipping a layer to account for geometrical acceptance (dashed arrow).}
    \label{fig:adjacentlayers}
\end{figure}

The doublets are created by connecting hits belonging to adjacent pairs of pixel detector layers, illustrated by the solid arrows in Fig.~\ref{fig:adjacentlayers}.
To account for geometrical and detector inefficiency doublets are also created between chosen pairs of non-adjacent layers, as illustrated by the dashed arrows in Fig.~\ref{fig:adjacentlayers}.

\begin{figure}[!tbp]
    \centering
    \includegraphics[width=0.7\textwidth]{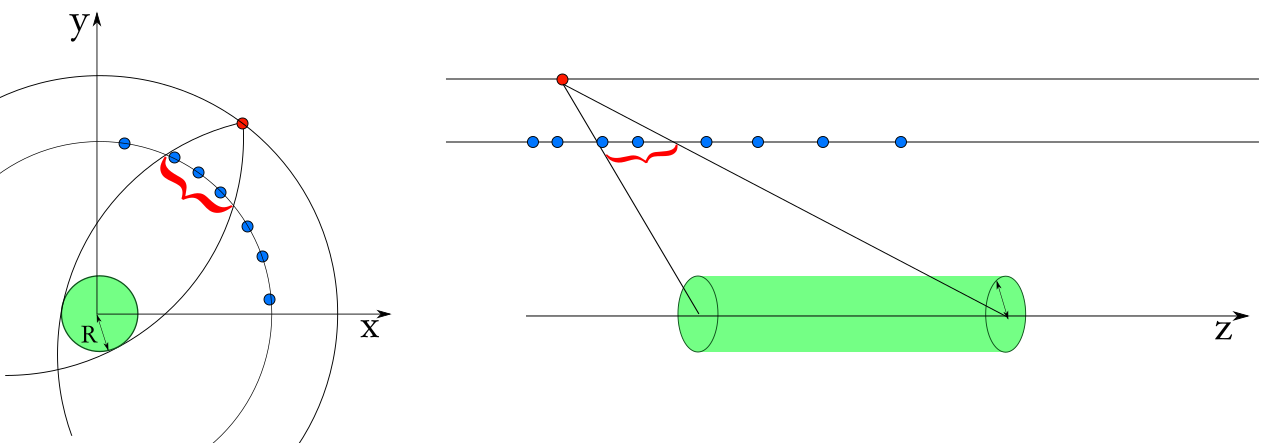}
    \caption{Windows opened in the transverse and longitudinal planes. The outer hit is colored in red, the inner hits in blue~\cite{Pantaleo:2293435}.
    }
    \label{fig:seedingwindow}
\end{figure}

Various selection criteria are applied to reduce the combinatorics. The following criteria have a strong impact on timing and physics performance:
\begin{itemize}
\item $p_\mathrm{T}^{\mathrm{min}}$: searching for low transverse momentum tracks can be very computationally expensive. Setting a minimum threshold for $p_\mathrm{T}$ limits the possible curvature, hence reducing the number of possible combinations of hits.
\item $R^{\mathrm{max}}$ and $z^{\mathrm{max}}$: the maximum transverse and longitudinal distance of closest approach with respect to the beam-spot. Tracks produced within a radius of less than $1$~mm around the beam-spot are called \emph{prompt tracks}. Searching for \emph{detached tracks} with a larger value of $R^{\mathrm{max}}$ leads to an increase in combinatorics. These ``alignment criteria'' are illustrated in Fig.~\ref{fig:seedingwindow}.
\item $n_{\mathrm{hits}}$: requiring a high number of hits in the \ntup{s} leads to a more pure set of tracks and cuts can be loosened, while a lower number of hits produces higher efficiency at the cost of a higher fake-rate.
\end{itemize}

Hits within each layers are arranged in a tiled data-structure along the azimuthal ($\phi$) direction for optimal performance. The search for compatible hit pairs is performed in parallel by different threads, each starting from a different outer hit. The pairs of inner and outer hits that satisfy the alignment criteria and have compatible clusters sizes along the z-direction form a doublet. The cuts applied during the doublets building are described in Table~\ref{tab:DoubletsCuts}, and their impact on the physics results and reconstruction time are provided in Tables~\ref{table:DoubletsPerformance} and~\ref{table:DoubletsTime}.

\begin{table}[!p]
    \centering
    \begin{tabular}{||l|p{9cm}||}
    \hline
         \textbf{Cut }& \textbf{Description} \\ \hline
 PhiHist & binned phi window between inner and outer hit using a 128 bin histogram \\
 PhiW & PhiHist + tuned phi window between inner and outer hit \\
 ZW & window in z for the inner hit \\
 ZIP & cut on the impact parameter along the beam axis \\
 PT & cut on the curvature assuming zero transverse impact parameter (equivalent to a cut on the TIP for high pt tracks) \\
 CZS & cut on the cluster size compatibility \\ 
 \hline
    \end{tabular}
    \caption{Description of the cuts applied during the reconstruction of doublets.}
    \label{tab:DoubletsCuts}
\end{table}

\begin{table}[!p]
\centering
 \begin{tabular}{||l | r| r| r| r||} 
 \hline
 Cuts & Doublets & \ntup{s} & Tracks & unconn \\ 
 \hline\hline
PhiHist & 1,268,193 &  23,254 & 1,256 & 0.966  \\
PhiHist+ZW  & 866,316 & 18,301 &  1,266 & 0.966 \\
PhiHist+ZW+ZIP & 269,410 & 11,235 &  1,265 &  0.926  \\
PhiW+ZW & 594,739 & 13,403 & 1,212 &  0.958  \\
PhiW+ZW+ZIP & 185,642 &  8,327 &  1,214 & 0.919  \\
PhiW+ZW+ZIP+CSZ & 129,307 &   6,060 & 1,087 &  0.915  \\
PhiW+ZW+ZIP+PT & 164,567 &  7,273 & 1,141 & 0.921  \\
PhiW+ZW+ZIP+PT+CSZ & 115,248 &  5,270 & 999 & 0.918 \\ [1ex] 
 \hline
 \end{tabular}
\caption{
Average number of doublets, \ntup{s} and final tracks per event, as well as the fraction of cell not connected, for each set of doublet reconstruction cuts (described in Table \ref{table:DoubletsPerformance}), running over a sample of $t\bar{t}$ events with an average pileup of $50$ and an average of 15,000 hits per event. 
}
\label{table:DoubletsPerformance}
\end{table}

\begin{table}[!p]
\centering
 \begin{tabular}{||l | r| r| r| r||} 
 \hline
 \multicolumn{1}{||c|}{} &  
 \multicolumn{4}{|c||}{time in $\mu$s} \\
 \hline
 Cuts &  doublets & connect & DFS & clean \\ 
 \hline\hline
PhiHist &   6,123 &  15,127 &  1,690 &  1,976 \\
PhiHist+ZW  &   950 &  6,582 &  778 &   538 \\
PhiHist+ZW+ZIP &  310 &  488 &   354 &  237 \\
PhiW+ZW & 552 & 2,995 &  549 &   377 \\
PhiW+ZW+ZIP &  271& 265 & 274 &  183 \\
PhiW+ZW+ZIP+CSZ &  291 & 187 & 216 &  154 \\
PhiW+ZW+ZIP+PT &  259 &156 & 246 &  125\\
PhiW+ZW+ZIP+PT+CSZ &  280 & 108 & 192 & 114\\  [1ex]
 \hline
 \end{tabular}
\caption{
Time spent in the three components of \ntup{s} building, as well as in the \emph{Fishbone} and ambiguity resolution algorithms (``clean''), for each set of doublet reconstruction cuts (described in Table \ref{table:DoubletsPerformance}).
It should be noted that using very relaxed cuts requires larger memory buffers on GPU, up to 12GB, while running with the last 4 sets requires less than 2GB of memory.
}
\label{table:DoubletsTime}
\end{table}

The doublets that share a common hit are tested for compatibility to form a triplet. The compatibility requires that the three hits are aligned in the $R-Z$ plane, and that the circumference passing through them intersects the beamspot compatibility region defined by $R_{\mathrm{max}}$. All doublets from all layer pairs are tested in parallel.

All compatible doublets form a direct acyclic graph. 
All the doublets whose inner hit lies on \emph{BPix1} are marked as root doublets. To reconstruct "outer" triplets, doublets starting on \emph{BPix2} or the two \emph{FPix1} layers and without inner neighbors are also marked as root.  
Each root doublet is subsequently assigned to a different thread that performs a Depth-First Search (DFS) over the direct acyclic graph starting from it. A DFS is used because one could prefer searching for all the \ntup{s} up to $n$ hits. The advantage of this approach is that the buckets containing triplets and quadruplets are disjoint sets as a triplet could not have been extended further to become a quadruplet~\cite{Pantaleo:2293435}.

\subsection{Fishbone \ntup{s}}

\begin{figure}[!htbp]
\centering
     \includegraphics[width=0.5\textwidth]{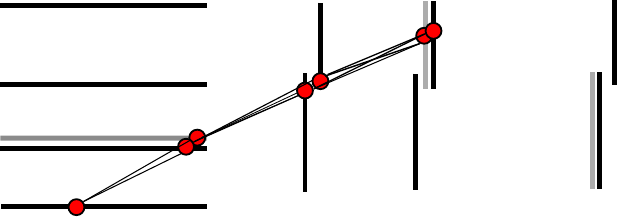}
     \caption{A typical \emph{Fishbone} \ntup{}. The shadowed areas indicate partially overlapping modules in the same layer.}
     \label{fig:fishbone}
\end{figure}

Full hit coverage in the instrumented pseudorapidity range is implemented in modern Pixel Detectors via partially overlapping sensitive layers. This, at the same time, mitigates the impact of possible localized hit inefficiencies. With this design, though, requiring at most one hit per layer can lead to several \ntup{s} corresponding to the same particle. This is particularly relevant in the forward region due to the design of the Pixel Forward Disks that is illustrated in Fig.~\ref{fig:fishbone}: up to four hits in the same layer can be found in localized forward areas. The \emph{Fishbone} \ntup{} solves the ambiguities by merging overlapping doublets. The \emph{Fishbone} mechanism is active while creating the doublets: among all the aligned doublets that share the same outermost hit, only the shortest one is kept. In this way ambiguities are resolved and a single \emph{Fishbone} \ntup{} is created.

Furthermore, among all the tracks that share a hit-doublet only the ones with the largest number of hits are retained.

\subsection{\ntup{} fit}

The ``Phase 1'' upgraded pixel detector has one more barrel layer and one additional disk at each side with respect to the previous detector. The possibility of using four (or more) hits from distinct layers opens new opportunities for the pixel tracker fitting method. It is possible not only to give a better statistical estimation of the track parameters ($\mathrm{d}_z$, $\mathrm{cot}\left(\theta\right)$, $\mathrm{d}_0$, $p_\mathrm{T}$ and $\phi$~\cite{fru2000data})  thanks to the additional point, but also to include in the fitting procedure more realistic effects, such as the energy loss and the multiple scattering of the particle due to its interaction with the material of the detector.

The pixel track reconstruction developed by the authors includes a multiple scattering-aware fit: the Broken Line~\cite{Blobel:2006yi} Fit. This follows three main steps:
\begin{itemize}
    \setlength\itemsep{0em}
    \item a fast pre-fit in the transverse plane gives an estimate of the track momentum, used to compute the multiple scattering contribution,
    \item a line-fit in the S-Z plane,
    \item a circle fit in the transverse plane.
\end{itemize}

The $\mathrm{d}_z$ and $\mathrm{cot}\left(\theta\right)$ track parameters and their covariance matrices are derived from the line fit, while the $\mathrm{d}_0$, $p_\mathrm{T}$ and $\phi$, and their covariance matrices from the circle fit. The final track parameters and covariance matrix are computed combining the individual results together.

The fits are performed in parallel over all \ntup{s} using one thread per \ntup{}. The fit implementation uses the Eigen C++ library~\cite{Eigen} that natively supports CUDA.

\subsection{Ambiguity resolution}
Tracks that share a hit-doublet are considered ``ambiguous'' and only the one with the best $\chi^2$ is retained. Triplets are considered ``ambiguous'' if they share one hit: only the one with the smallest transverse impact parameter is retained.

\subsection{Pixel Vertices}

The fitted pixel tracks are subsequently used to form pixel vertices.
Vertices are searched as clusters in the $z$ coordinate of the point of closest transverse approach of each track with the beam line ($z_0$).
Only tracks with at least 4 hits and a $p_\mathrm{T}$ larger than a configurable threshold (0.5 GeV) are considered. For each track with an error in $z_0$ lower than a configurable threshold the local density of close-by tracks is computed. Tracks are considered in the density calculation if they are within a certain $\Delta z_{\mathrm{cut}}$ and if their $\chi^2$ compatibility is lower than a configurable $\chi^2_\mathrm{threshold}$. Tracks with local density greater than $1$ are considered as a seed for a vertex. Tracks are then linked to another track that has a higher local density, if the distance between the two tracks is lower than $\Delta z_{\mathrm{cut}}$ and if their $\chi^2 \leq \chi^2_\mathrm{threshold}$. All the tracks that are logically linked starting from each seed become part of the same vertex candidate. Each vertex candidate is promoted to be a final vertex if it contains at least 2 tracks.

This algorithm is easily parallelizable and, in one dimension as in this case, requires no iterations. It is less sensitive to noise (fake tracks) and has a lower merge rate of a standard DBSCAN~\cite{conf/kdd/EsterKSX96}.
It is much faster than any hierarchical algorithm~\cite{HDBSCAN} or algorithms based on deterministic annealing~\cite[\S6.1]{Chatrchyan:2014fea}.
As showed below this algorithm is definitively more efficient and has comparable resolution than the "gap" algorithm used so far at the CMS HLT~\cite[\S6.2]{Chatrchyan:2014fea}.

Each vertex position and error along the beam line are computed from the weighted average of the $z_0$ of the contributing tracks. Vertices with a $\chi^2$ larger than a given threshold (9 per degree of freedom) are split in two using a {\it k-mean} algorithm.

Finally the vertices are sorted using the sum of the  $p_\mathrm{T}^2$ of the contributing tracks. The vertex with the largest $\sum p_\mathrm{T}^2$ is labelled as the ``primary'' vertex i.e. the vertex corresponding to the signal (triggering) event.

\section{Results}\label{sec:results}

In this section the performance of the Patatrack reconstruction is evaluated and compared to the track reconstruction that CMS has used for data taking in 2018 (in the following referred to as CMS-2018)~\cite{Pantaleo:2293435}.

\subsection{Input dataset}
The performance studies have been performed using \num{20000} t$\bar{\mathrm{t}}$ simulated events from CMS open data~\cite{cms2019opendata}, with an average of \num{50} superimposed pileup collisions with a center-of-mass energy $\sqrt{s}=13$~TeV, using detector design conditions. 

\subsection{Physics performance}
\begin{figure}[p]
\subfloat[Efficiency vs $p_\mathrm{T}$]{%
\includegraphics[width=0.45\textwidth]{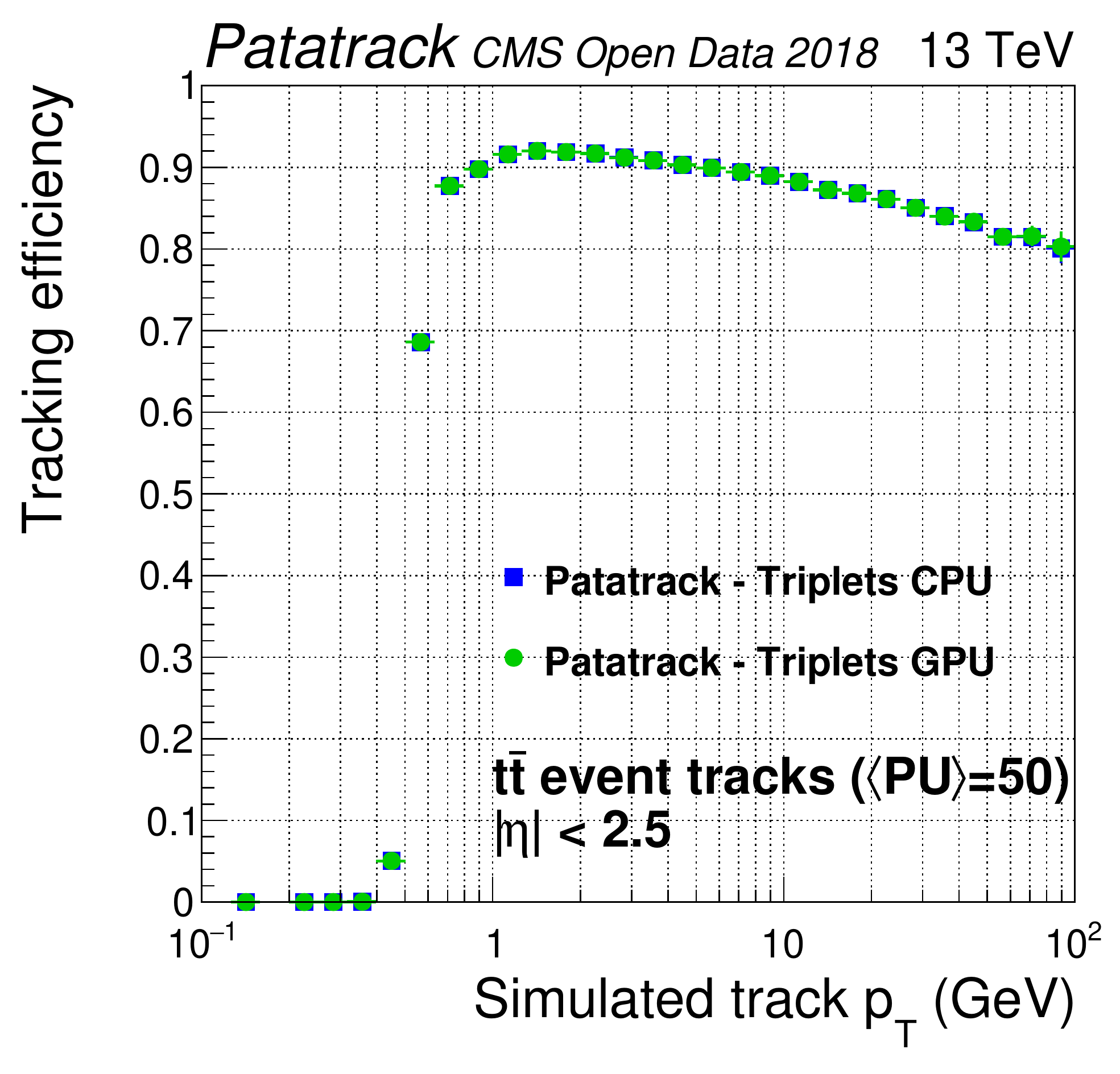}
}
\hfill
\subfloat[Efficiency vs $\eta$]{%
\includegraphics[width=0.45\textwidth]{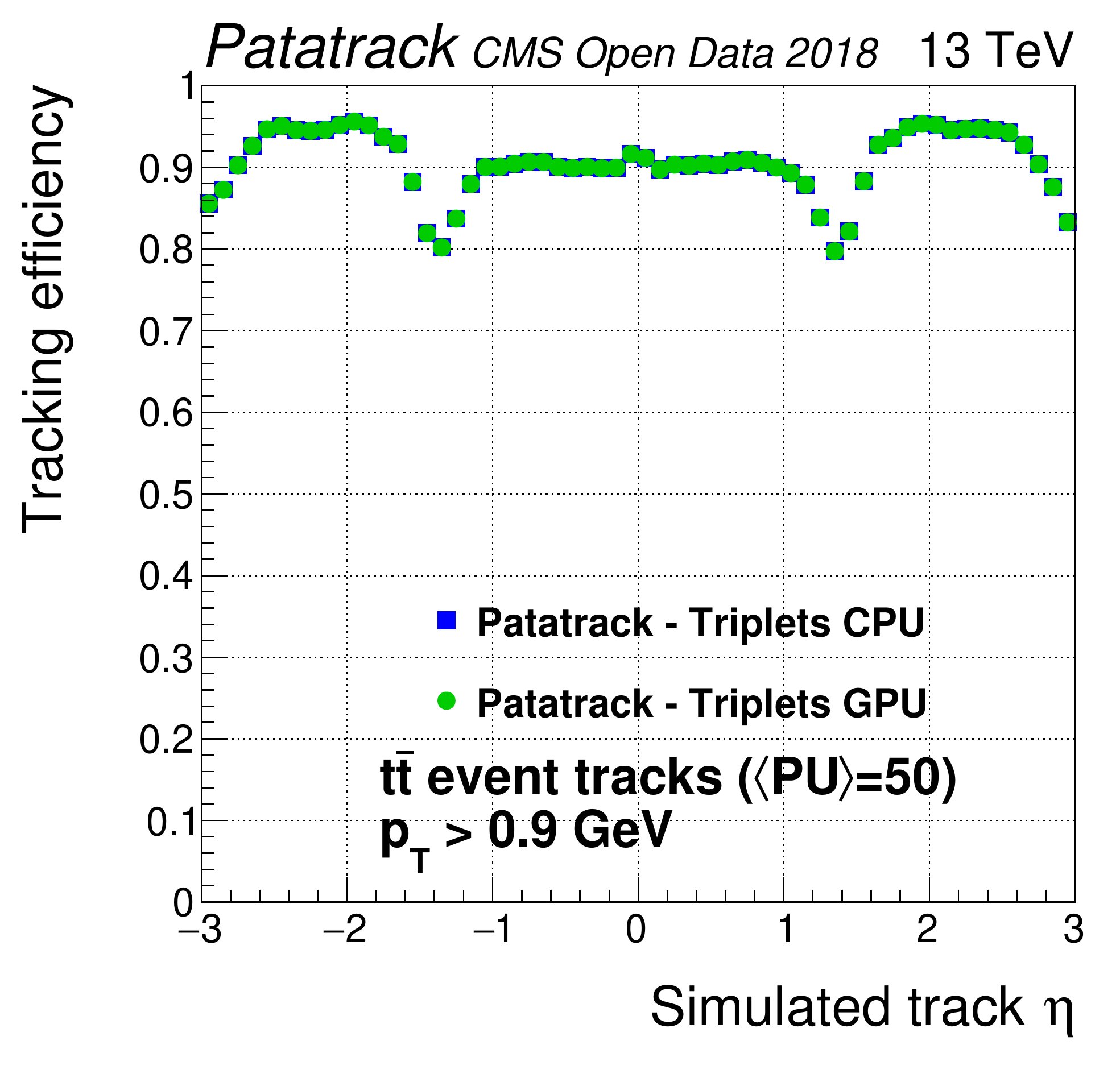}
}
\caption{Comparison of the pixel tracks reconstruction efficiency of the CPU and GPU versions of the Patatrack Pixel reconstruction for simulated t$\bar{\mathrm{t}}$ events with an average of 50 superimposed pileup collisions.}
\label{fig:efficiency_CPU_GPU}
\end{figure}

The efficiency is defined as the fraction of simulated tracks, $N_{sim}$, having produced at least three hits in the pixel detector, that have been associated with at least one reconstructed track, $N_{rec}$.
\begin{equation}
\mathrm{efficiency}=\frac{N_{rec}}{N_{sim}}
\end{equation}
A reconstructed pixel track is associated with a simulated track if all the hits that it contains come from the same simulated track. The efficiency is computed only for tracks coming from the hard interaction and not for those from the pileup.
The CPU and GPU versions of the Patatrack workflow produce the same physics results, as shown in Fig.~\ref{fig:efficiency_CPU_GPU}. For this reason, there will be no further distinction in the discussion of the physics results between the workflows running on CPU and GPU.

\begin{figure}[p]
\subfloat[Efficiency vs $p_\mathrm{T}$]{%
\includegraphics[width=0.45\textwidth]{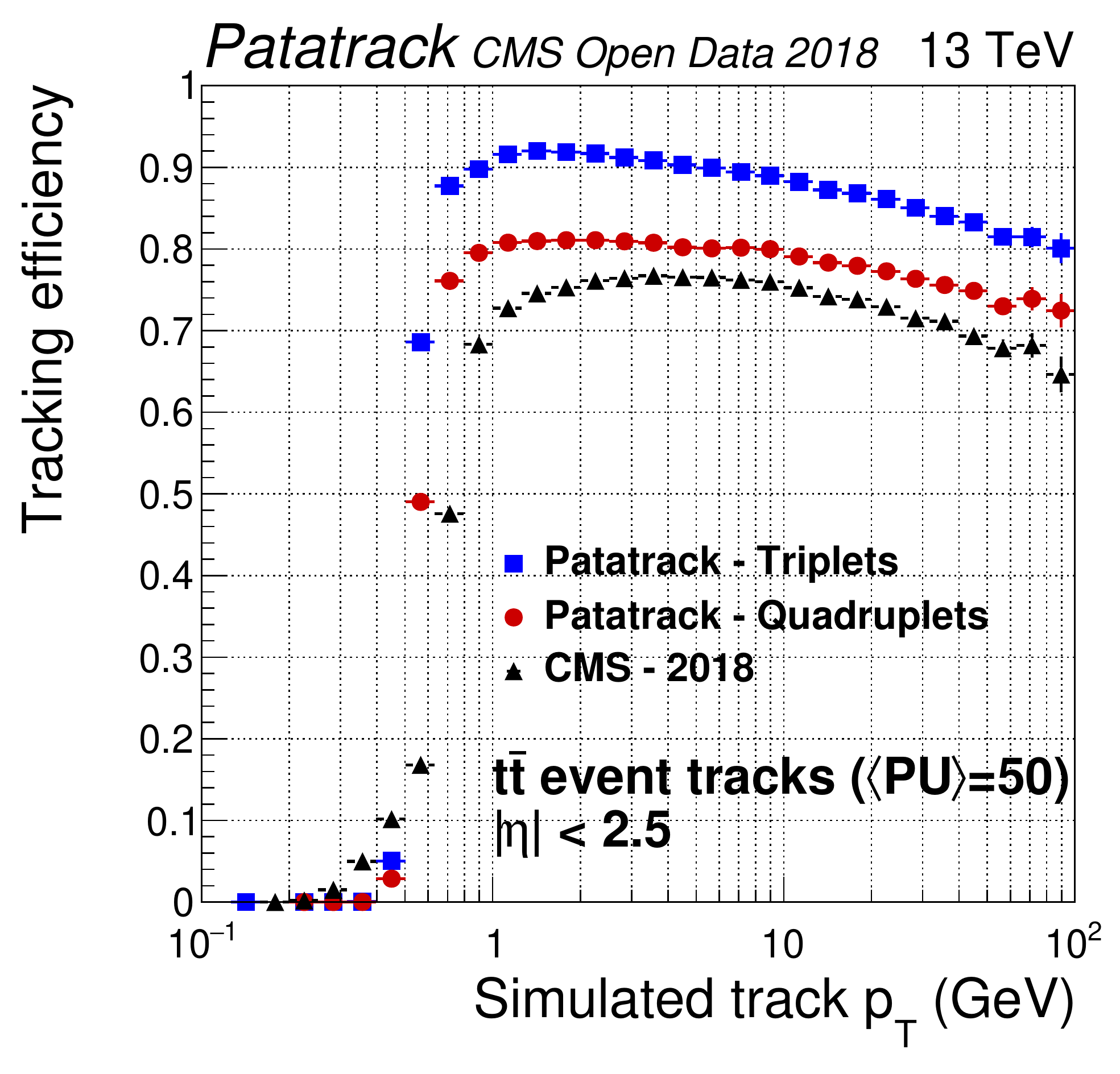}
}
\hfill
\subfloat[Efficiency vs $\eta$]{%
\includegraphics[width=0.45\textwidth]{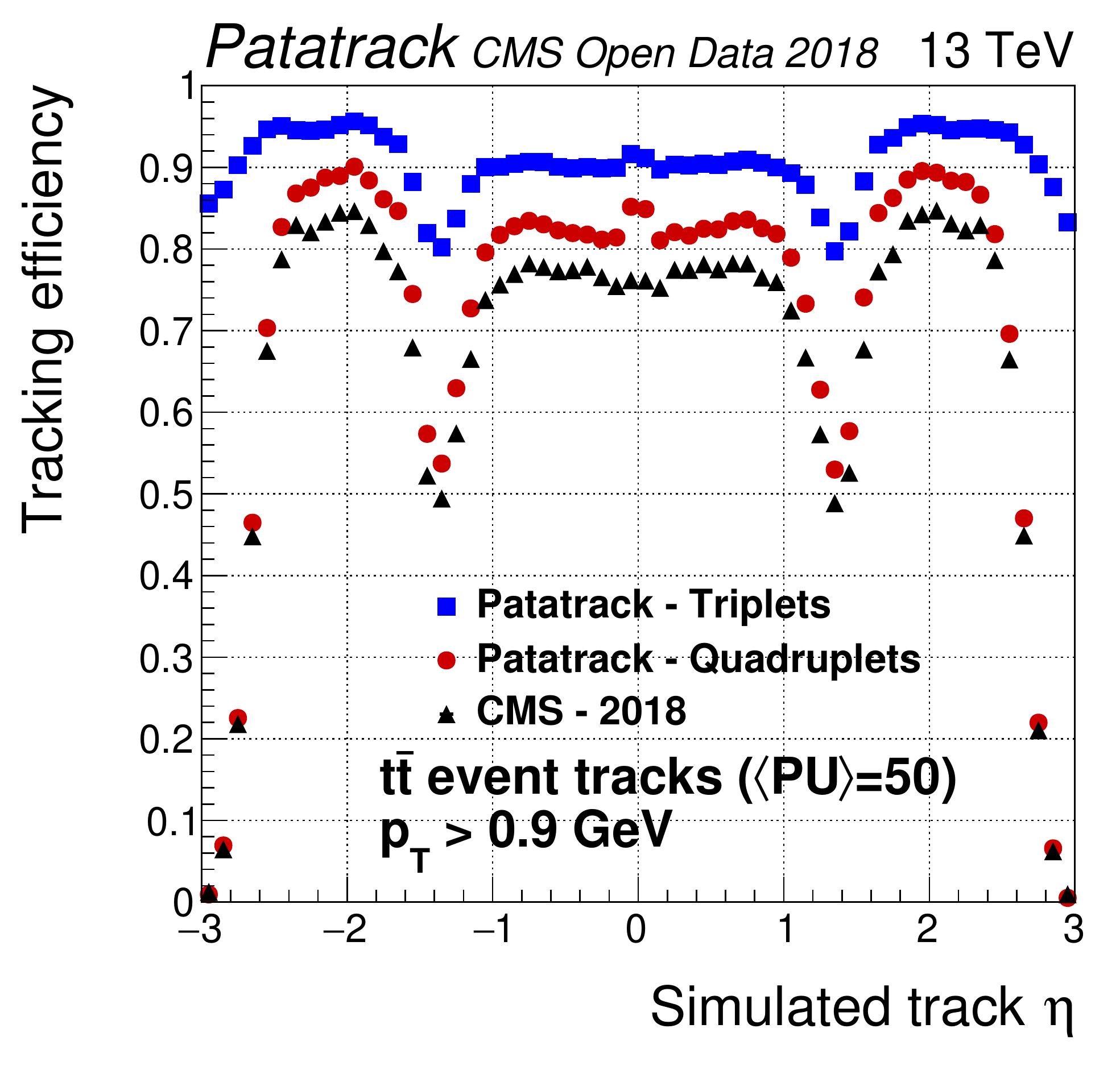}
}
\caption{Pixel tracks reconstruction efficiency for simulated t$\bar{\mathrm{t}}$ events with an average of 50 superimposed pileup collisions. The performance of the Patatrack reconstruction when producing Pixel Tracks starting from \ntup{s} with $n_{hits}\geq 3$ and $n_{hits}\geq 4$ are represented respectively by blue squares and red circles. The performance of CMS-2018 is represented by black triangles.}
\label{fig:efficiency_ttbar}
\end{figure}

The efficiency of quadruplets is sensibly improved by the Patatrack quadruplets workflow with respect to CMS-2018, as shown in Fig.~\ref{fig:efficiency_ttbar}. The main reasons for this improvement are the possibility to skip a layer outside geometrical acceptance when building doublets and the usage of different Cellular Automaton cuts for the barrel and the end-caps. The efficiency can be further improved including the pixel tracks built from triplets (Patatrack triplets).

\begin{figure}[p]
\subfloat[Fake rate vs $p_\mathrm{T}$]{%
\includegraphics[width=0.45\textwidth]{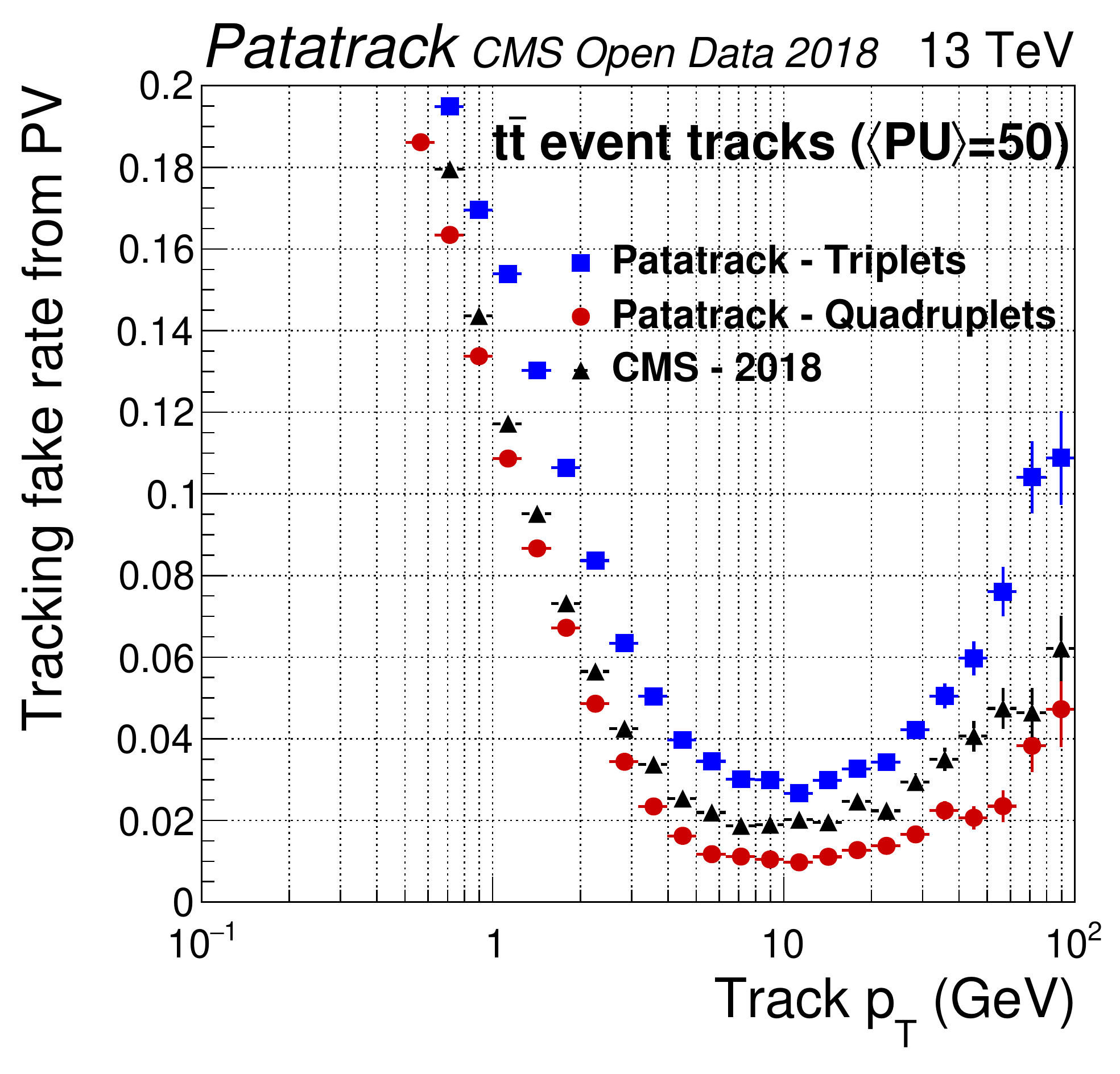}
}
\hfill
\subfloat[Fake rate vs $\eta$]{%
\includegraphics[width=0.45\textwidth]{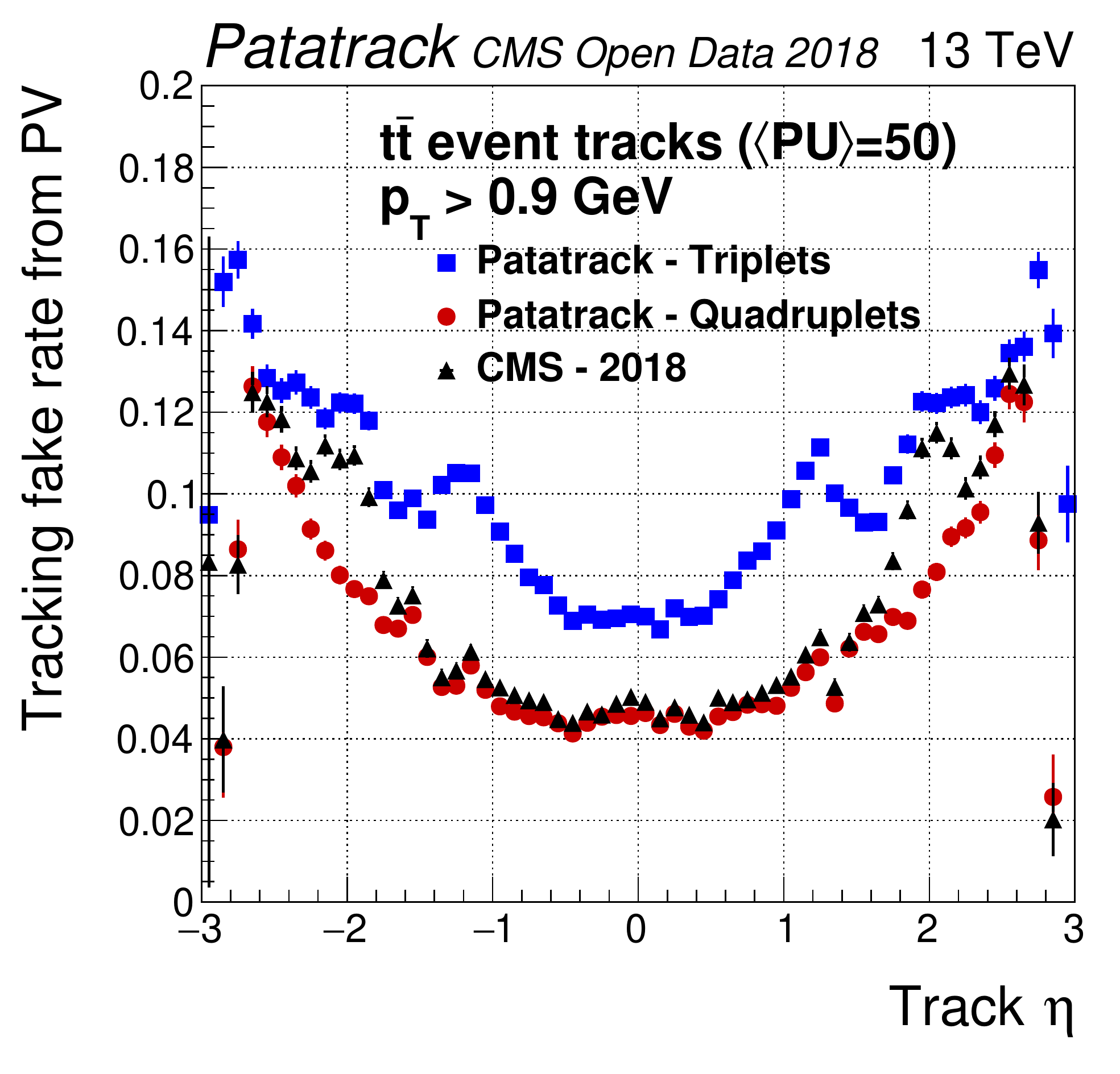}
}
\caption{Pixel tracks reconstruction fake rate for simulated t$\bar{\mathrm{t}}$ events with an average of 50 superimposed pileup collisions. The performance of the Patatrack reconstruction when producing pixel tracks starting from \ntup{s} with $n_{hits}\geq 3$ and $n_{hits}\geq 4$ are represented respectively by blue squares and red circles. The performance of CMS-2018 is represented by black triangles.}
\label{fig:fakerate_ttbar}
\end{figure}
The fake rate is defined as the fraction of all the reconstructed tracks coming from a reconstructed primary vertex that are not associated uniquely to a simulated track. In the case of a fake track, the set of hits used to reconstruct the track does not belong to the same simulated track. As shown in Fig.~\ref{fig:fakerate_ttbar}, the fake-rate performance of Patatrack quadruplets is improved with respect to the CMS-2018 pixel reconstruction in the end-cap region, mainly thanks to the different treatment of the end-caps in the Cellular Automaton. The inclusion of the pixel tracks built from Patatrack triplets slightly increases the fake rate in the tracks coming from the primary vertices,  given that loosening the requirement on the number of hits decreases the quality of the selection cuts.

\begin{figure}[p]
\subfloat[Duplicate rate vs $p_\mathrm{T}$]{%
\includegraphics[width=0.45\textwidth]{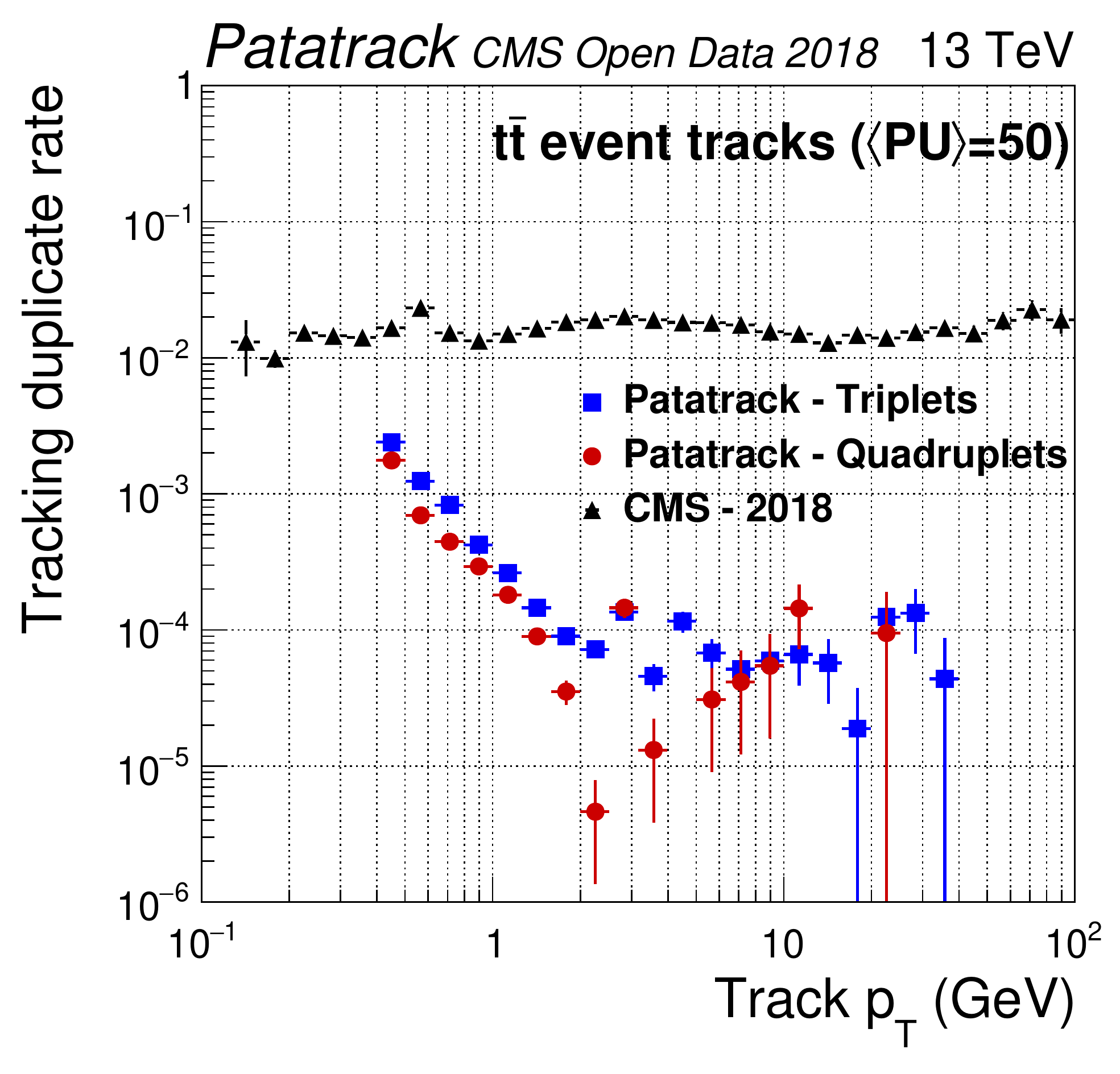}
}
\hfill
\subfloat[Duplicate rate vs $\eta$]{%
\includegraphics[width=0.45\textwidth]{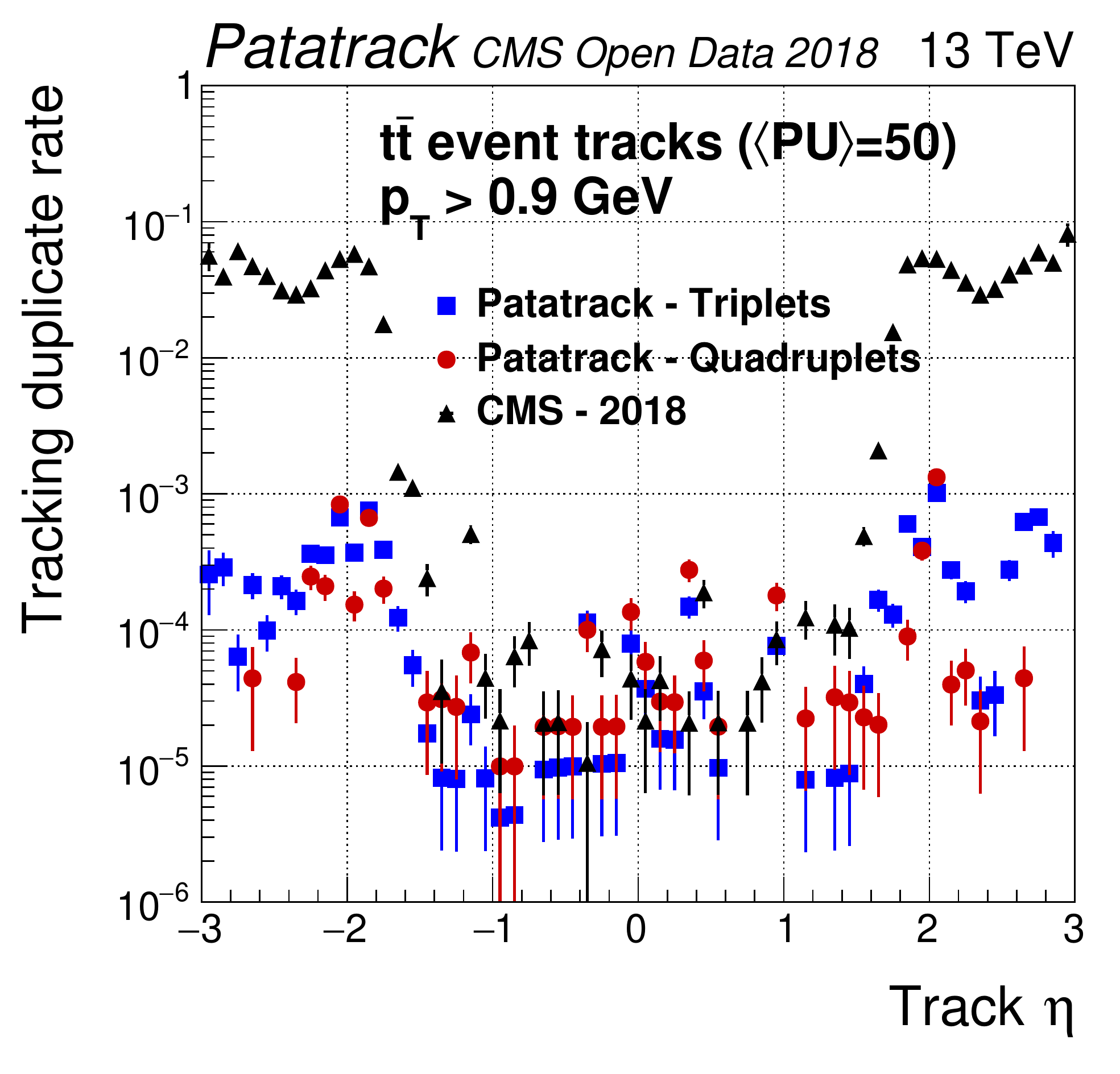}
}
\caption{Pixel tracks reconstruction duplicate rate for simulated t$\bar{\mathrm{t}}$ events with an average of 50 superimposed pileup collisions. The performance of the Patatrack reconstruction when producing Pixel Tracks starting from \ntup{s} with $n_{hits}\geq 3$ and $n_{hits}\geq 4$ are represented respectively by blue squares and red circles. The performance of CMS-2018 is represented by black triangles.}
\label{fig:duplicaterate_ttbar}
\end{figure}
A duplicate track is a reconstructed track matching to a simulated track that itself has been matched to at least two tracks. The introduction of the Fishbone algorithm improves the duplicate rejection in the Patatrack workflows by up to two orders of magnitude with respect to the CMS-2018 pixel track reconstruction, as shown in Fig.~\ref{fig:duplicaterate_ttbar}.

\begin{figure}[tb]
\subfloat[$p_\mathrm{T}$ resolution vs $p_\mathrm{T}$]{%
\includegraphics[width=0.45\textwidth]{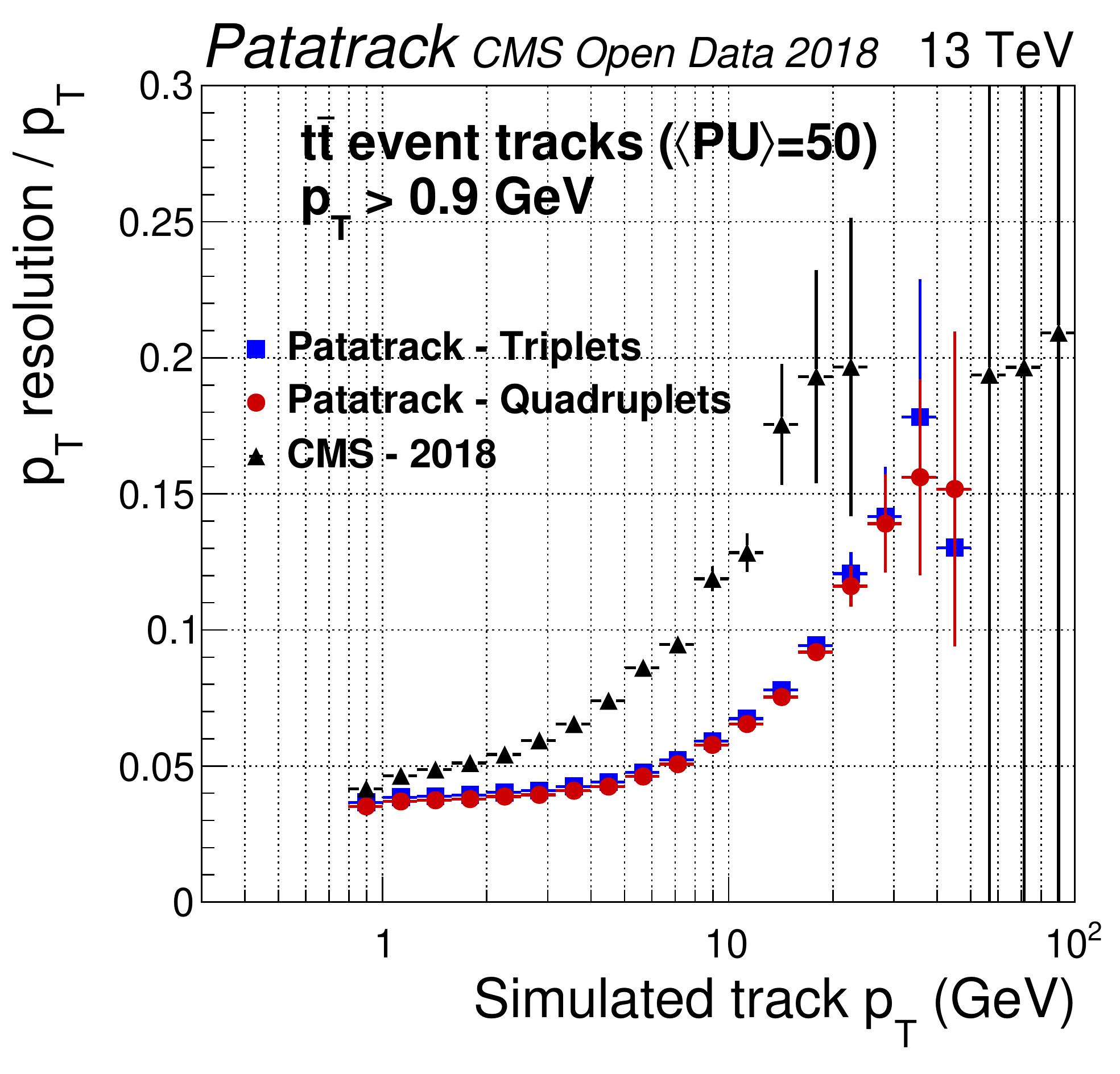}
}
\hfill
\subfloat[$p_\mathrm{T}$ resolution vs $\eta$]{%
\includegraphics[width=0.45\textwidth]{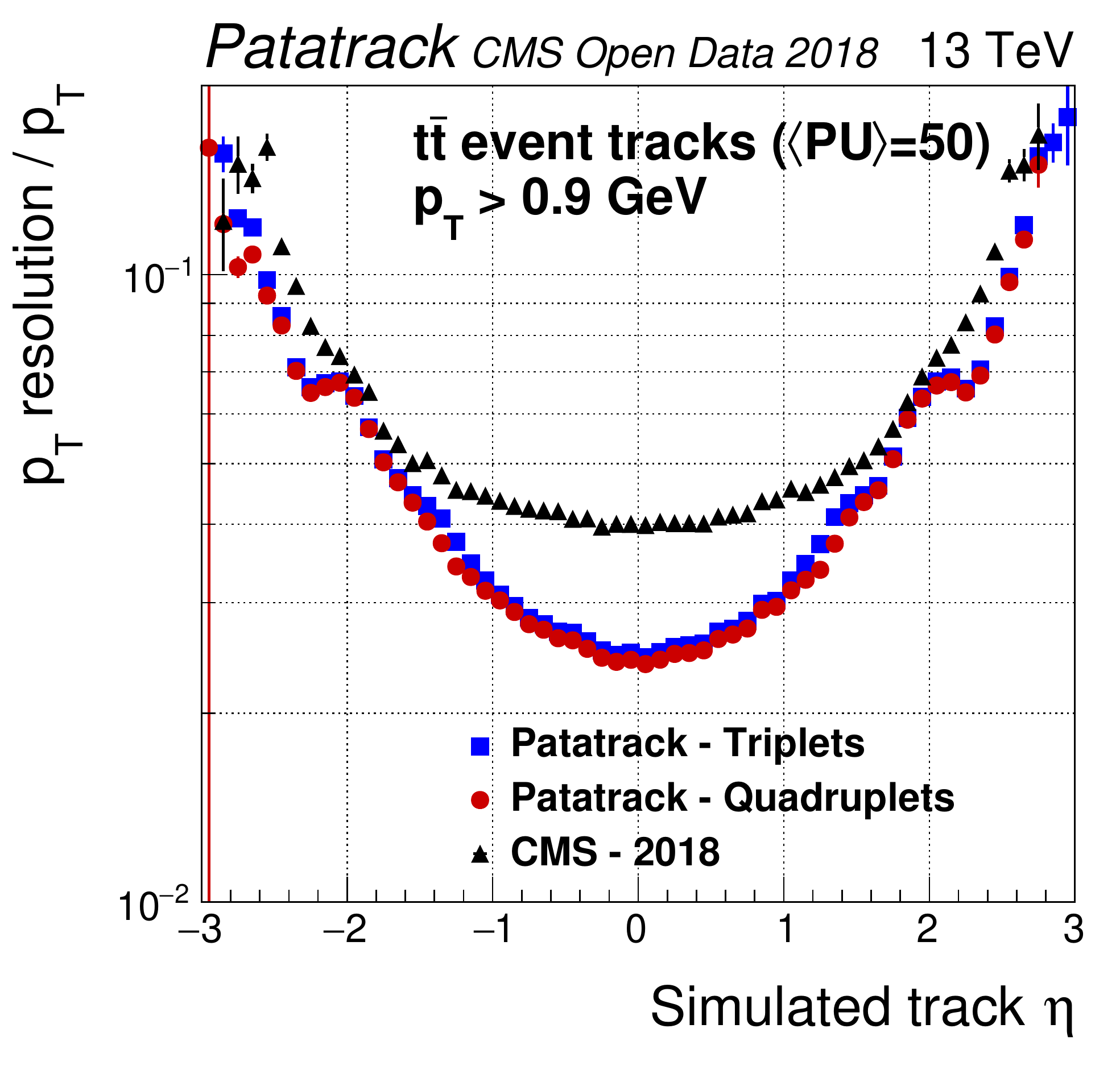}
}
\caption{Pixel tracks $p_\mathrm{T}$ resolution for simulated t$\bar{\mathrm{t}}$ events with an average of 50 superimposed pileup collisions. The performance of the Patatrack reconstruction when producing Pixel Tracks starting from \ntup{s} with $n_{hits}\geq 3$ and $n_{hits}\geq 4$ are represented respectively by blue squares and red circles. The performance of CMS-2018 is represented by black triangles.}
\label{fig:pT_resolution_ttbar}
\end{figure}
For historical reasons the CMS-2018 Pixel reconstruction does not perform a fit on the \ntup{s} in the transverse plane, and considers instead only the first three hits for the track parameters estimation. Furthermore, the errors on the track parameters are taken from a look-up table parameterized in $\eta$ and $p_\mathrm{T}$. The improvement brought in  by the Broken Line fit to the accuracy of the fits can be quantified by looking at the resolutions defined as : 
\begin{equation}
    \sigma(\mathrm{fitted\;value - true \;value}).
\end{equation}

\begin{figure}[p]
\subfloat[$d_{xy}$ resolution vs $p_\mathrm{T}$]{%
\includegraphics[width=0.45\textwidth]{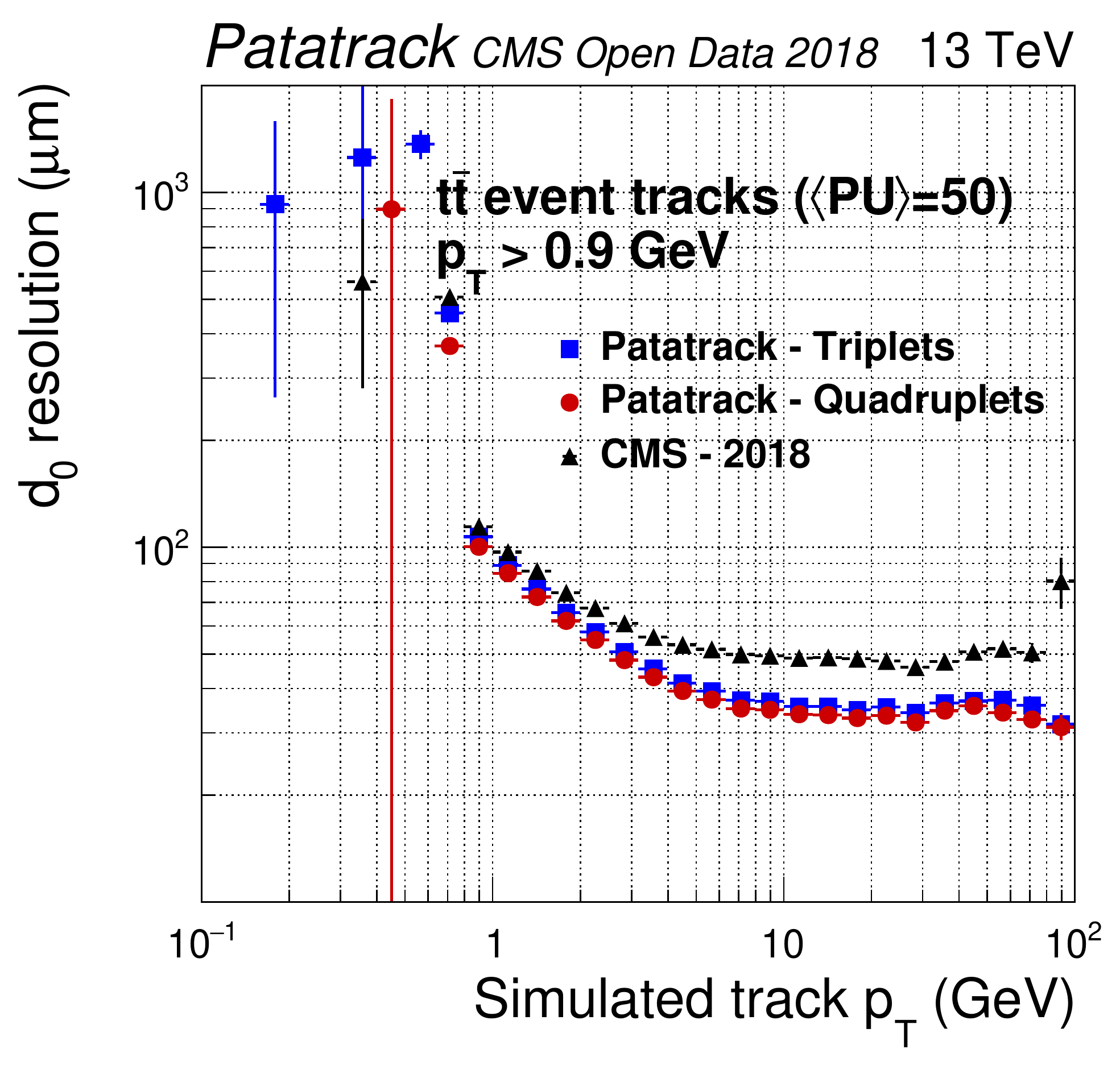}
}
\hfill
\subfloat[$d_{xy}$ resolution vs $\eta$]{%
\includegraphics[width=0.45\textwidth]{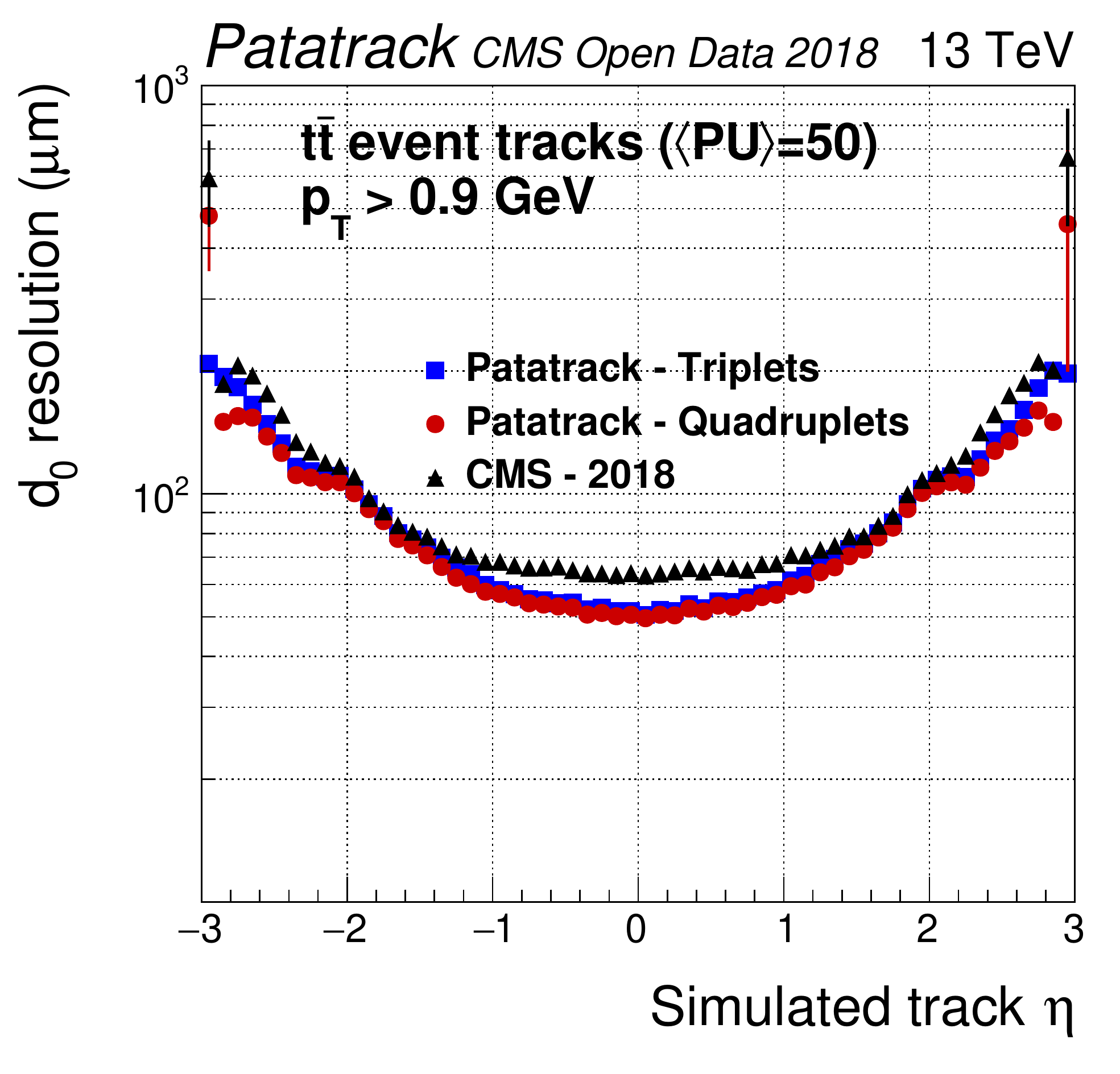}
}
\caption{Pixel tracks transverse impact parameter resolution for simulated t$\bar{\mathrm{t}}$ events with an average of 50 superimposed pileup collisions. The performance of the Patatrack reconstruction when producing pixel tracks starting from \ntup{s} with $n_{hits}\geq 3$ and $n_{hits}\geq 4$ are represented respectively by blue squares and red circles. The performance of CMS-2018 is represented by black triangles.}
\label{fig:dxy_resolution_ttbar}
\end{figure}
The resolution of the estimation of the $p_\mathrm{T}$ is improved by up to a factor $2$ when compared to the CMS-2018 Pixel tracking (Fig.~\ref{fig:pT_resolution_ttbar}).
The resolution of the transverse impact parameter $d_0$ improves, especially in the barrel (Fig.~\ref{fig:dxy_resolution_ttbar}).

\begin{figure}[p]
\subfloat[$d_z$ resolution vs $p_\mathrm{T}$]{%
\includegraphics[width=0.45\textwidth]{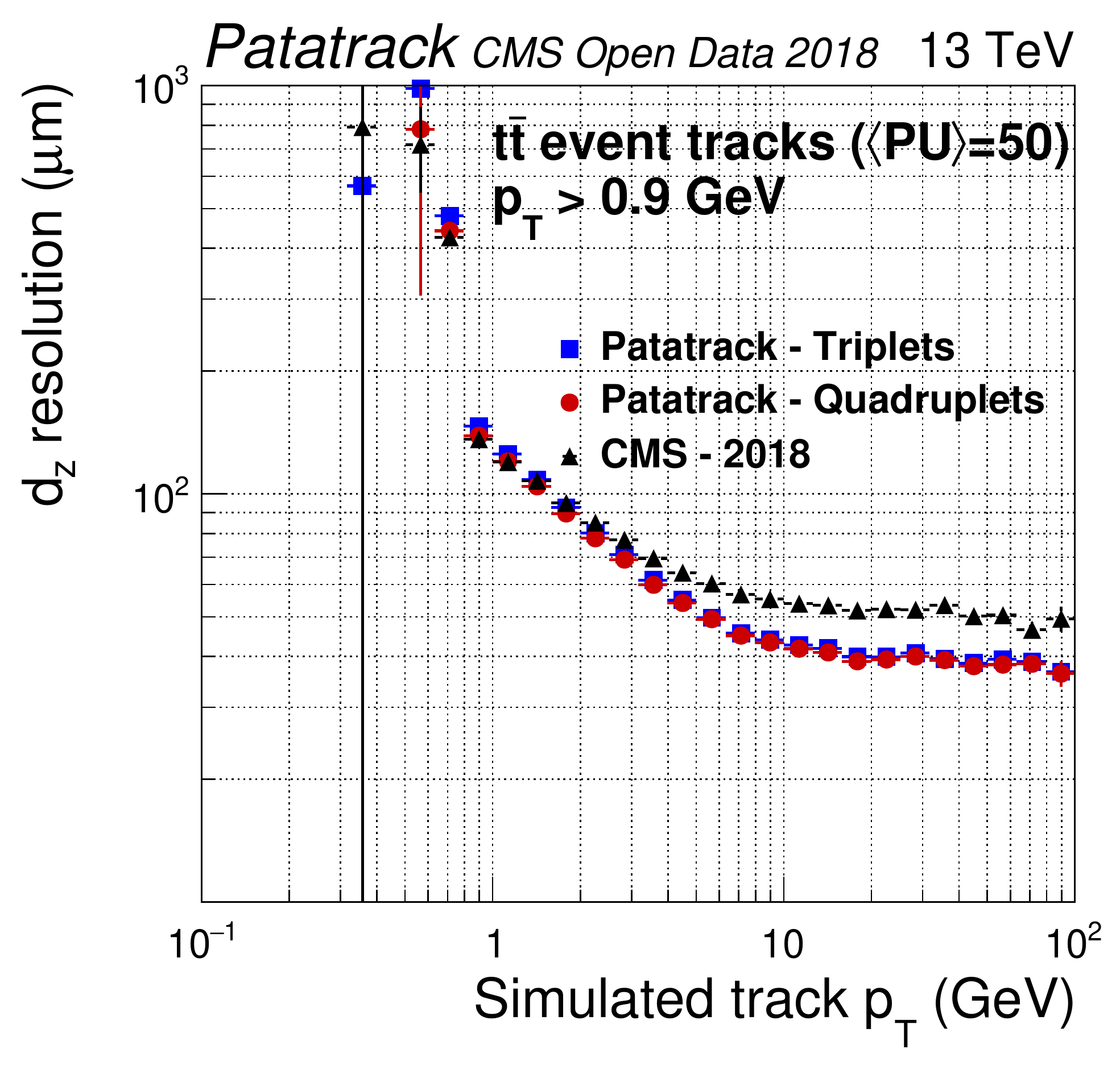}
}
\hfill
\subfloat[$d_z$ resolution vs $\eta$]{%
\includegraphics[width=0.45\textwidth]{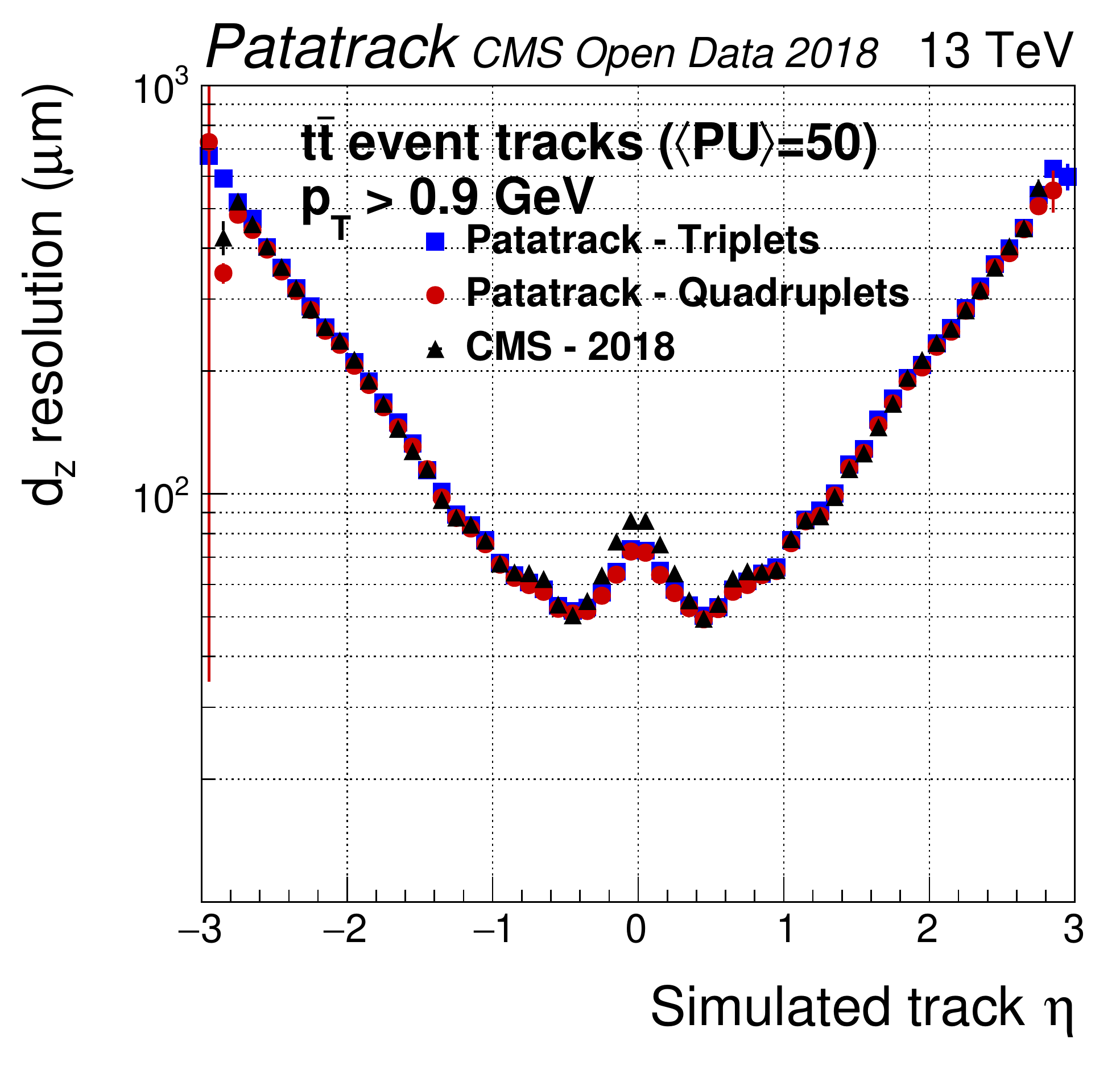}
}
\caption{Pixel tracks longitudinal impact parameter resolution for simulated t$\bar{\mathrm{t}}$ events with an average of 50 superimposed pileup collisions. The performance of the Patatrack reconstruction when producing pixel tracks starting from \ntup{s} with $n_{hits}\geq 3$ and $n_{hits}\geq 4$ are represented respectively by blue squares and red circles. The performance of CMS-2018 is represented by black triangles.}
\label{fig:dz_resolution_ttbar}
\end{figure}
The CMS 2018 Pixel tracking behaves better in the longitudinal plane than it does in the transverse plane. However, the Broken Line fit's improvement in the estimate of the longitudinal impact parameter $d_z$ is visible for tracks with $p_\mathrm{T}>3$~ $\mathrm{GeV}/c$, as shown by  (Fig.~\ref{fig:dz_resolution_ttbar}).

\begin{figure}[tb]
\subfloat[Reconstructed vs simulated vertices]{%
\includegraphics[width=0.45\textwidth]{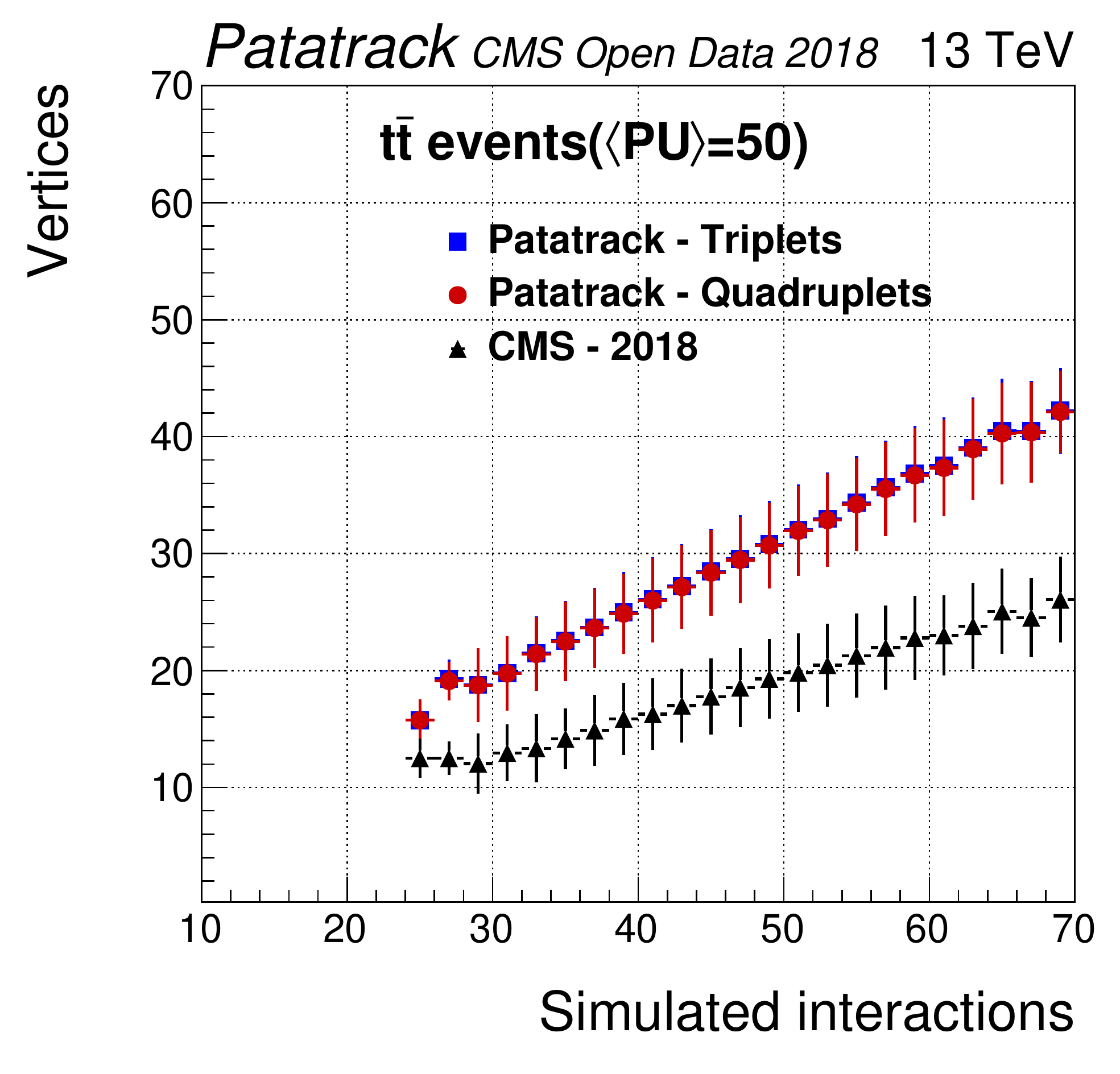}
}
\hfill
\subfloat[Vertex merge rate vs simulated vertices]{%
\includegraphics[width=0.45\textwidth]{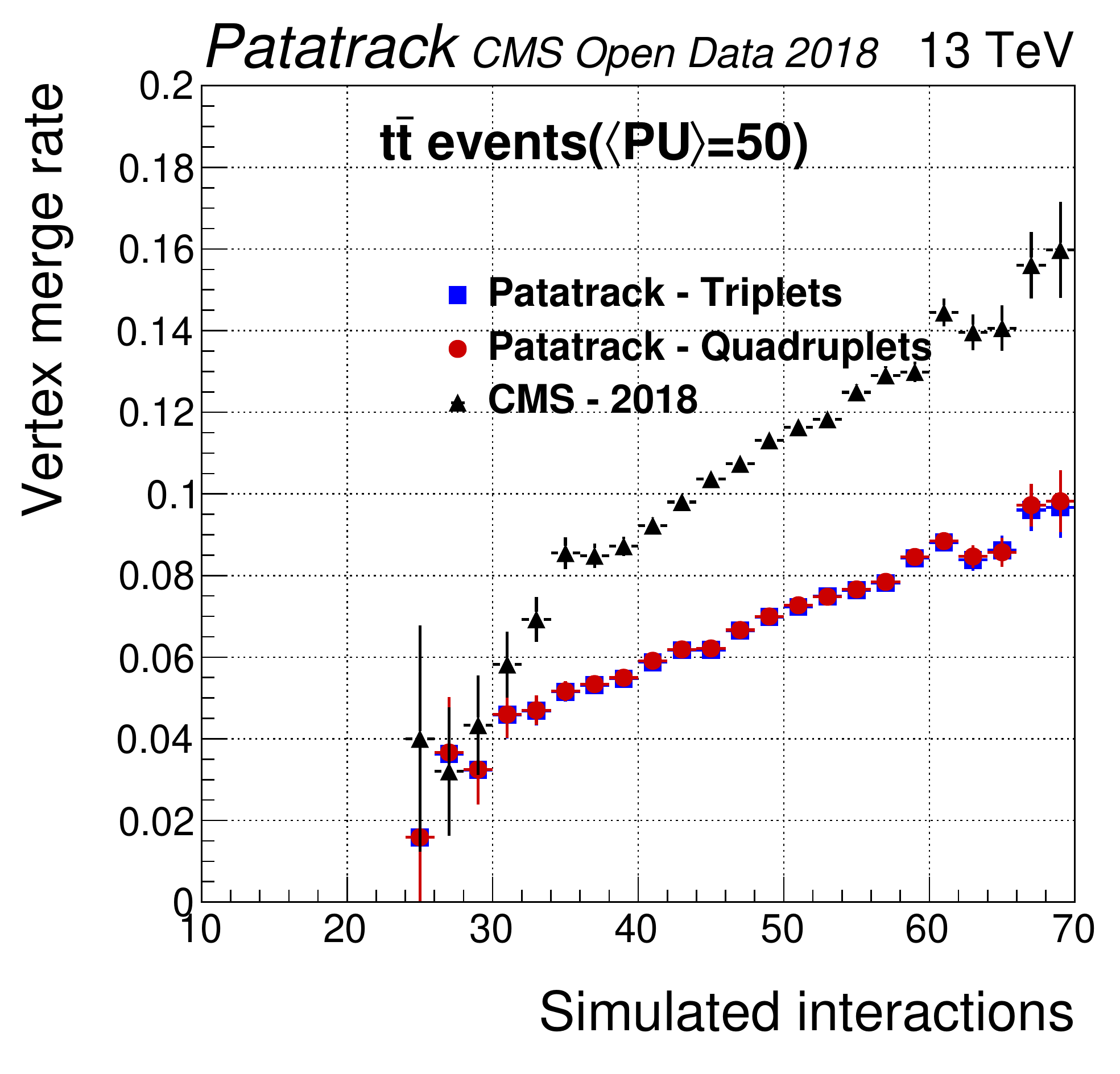}
}
\caption{Pixel vertices reconstruction efficiency and merge rate for simulated t$\bar{\mathrm{t}}$ events with an average of 50 superimposed pileup collisions. The performance of the Patatrack reconstruction when producing pixel tracks starting from \ntup{s} with $n_{hits}\geq 3$ and $n_{hits}\geq 4$ are represented respectively by blue squares and red circles. The performance of CMS-2018 is represented by black triangles.}
\label{fig:vertexing_ttbar}
\end{figure}

The number of reconstructed vertices together with the capability to separate two close-by vertices have been measured to have an estimate of the performance of the vertexing algorithm. This capability can be quantified by measuring the vertex merge rate, \emph{i.e.} the probability of having two different simulated vertices reconstructed as a single vertex.\\
Figure~\ref{fig:vertexing_ttbar} shows how the vertexing performance evolves with the number of simulated proton interactions.

\subsection{Computing performance}

The hardware and software configurations used to to carry out the computing performance measurements are:
\begin{itemize}
\setlength\itemsep{0em}
\item dual socket Xeon Gold 6130~\cite{intelXeon6130}, $2\times16$ physical cores, 64 hardware threads,
\item a single NVIDIA T4~\cite{nvidiat4},
\item NVIDIA CUDA 11 with Multi-Process Service~\cite{nvidiamps},
\item CMSSW 11\_1\_2\_Patatrack~\cite{cmsswpatatrack}.
\end{itemize}

A CMSSW reconstruction sequence that runs only the pixel reconstruction modules as described in Section~\ref{section:PatatrackSequence} was created. More than one event can be executed in parallel by different CPU threads. These can perform asynchronous operations like kernels and memory transfers, in parallel on the same GPU. The maximum amount of events that can be processed in parallel today is about 80, and is limited by the amount of allocated memory on the GPU required for each event. 

In a data streaming application the measurement of the throughput, i.e. the number of reconstructed events per unit time, is a more representative metric than the measurement of the latency.
The benchmark runs 8 independent CMSSW jobs, each reconstructing 8 events in parallel with 8 CPU threads.
The throughput of the CMS-2018 reconstruction has been compared to the Patatrack quadruplets and triplets workflows. The test includes the GPU and the CPU versions of the Patatrack workflows.
The Patatrack workflows run with three different configurations:
\begin{enumerate}
    \item \emph{no copy}: the SoA containing the results stays in the memory where they have been produced;
    \item \emph{copy, no conversion}: the SoA containing the results is copied to the host, if initially produced by the GPU;
    \item \emph{copy, conversion}: the SoA containing the results is copied to the host and converted to the legacy CMS-2018 pixel tracks and vertices data formats.
\end{enumerate}
These configurations are useful to understand the impact of optimizing a potential consumer of the GPU results so that it runs on GPUs in the same reconstruction sequence or so that it can consume GPU-friendly data structures, with respect to interfacing the Patatrack workflows to the existing framework without any further optimization.

\begin{table}[tb]
\centering
 \small{
 \begin{tabular}{||l | r| r| r| r| r||} 
 \hline
 \multicolumn{1}{||c|}{} &
 \multicolumn{5}{|c||}{throughput in events/s} \\
 \hline
 Configuration & Triplets CPU & Triplets GPU & Quadruplets CPU & Quadruplets GPU & CMS 2018  \\ 
 \hline
no copy        &  611 &  870 &  892 & 1386 &  476  \\
copy, no conv. &  --- &  867 &  --- & 1372 &  ---  \\
conversion     &  585 &  861 &  855 & 1352 &  ---  \\ 
 \hline\hline
 \end{tabular}
 }
\caption{
    Throughput of the Patatrack triplets and quadruplets workflows when executed on GPU and CPU, compared to the CMS-2018 reconstruction. The benchmark is configured to reconstruct 64 events in parallel. Three different configurations have been compared: in \emph{no copy} the result is not copied from the memory of the device where it was initially produced; in \emph{copy, no conv.} the SoA containing result produced on the GPU is copied to the host memory; in \emph{conversion} the SoA containing the result is copied to the host memory (if needed) and then converted to the legacy data format used for the pixel tracks and vertices by the CMS reconstruction.}
\label{table:ThroughputPerformance}
\end{table}

The results of the benchmark are shown in Table~\ref{table:ThroughputPerformance}.
The benchmark shows that a single NVIDIA T4 can achieve almost three times the performance of a full dual socket Intel Xeon Skylake node when running the Patatrack pixel quadruplets reconstruction. Producing even better physics performance by producing also pixel tracks from triplets has the effect of almost halving the throughput. 
Copying the results from the GPU memory to the host memory has a small impact to the throughput, thanks to the possibility of hiding latency by overlapping the execution of kernels with copies.
Converting the SoA results to the legacy data format has a small impact on the throughput as well, but comes with a hidden cost: the conversion takes almost 100\% of the machine's processing power. This can be avoided by migrating all the consumers to the SoA data format.


\section{Conclusions and future work}\label{sec:conclusion}
The future runs of the Large Hadron Collider (LHC) at CERN will pose significant challenges on the event reconstruction software, due to the increase in both event rate and complexity. For track reconstruction algorithms, the number of combinations that have to be tested does not scale linearly with the number of simultaneous proton collisions.

The work described in this article presents innovative ways to solve the problem of tracking in a pixel detector such as the CMS one, by making use of heterogeneous computing systems in a data taking production-like environment, while being integrated in the CMS experimental software framework CMSSW.
The assessment of the Patatrack reconstruction physics and timing performance demonstrated that it can improve physics performance while being significantly faster than the existing implementation. The possibility to configure the Patatrack reconstruction workflow to run on CPU or to transfer and convert results to use the CMS data format allows to run and validate the workflow on conventional machines, without any dedicated resources.

This work is setting the foundations for the development of heterogeneous algorithms in HEP both from the algorithmic and from the framework scheduling points of view. Other parts of the reconstruction, e.g. calorimeters or Particle Flow, will be able to benefit from an algorithmic and data structure redesign to be able to run efficiently on GPUs.

The ability to run on other accelerators with a performance portable code is also being explored, to ease maintainability and test-ability of a single source.

\section{Acknowledgements}

We thank our colleagues of the CMS collaboration for publishing high quality simulated data under the open access policy. 
We also would like to thank the Patatrack students and alumni R.~Ribatti and  G.~Tomaselli for their hard work and dedication.
We thank the CERN openlab for providing a platform for discussion on heterogeneous computing and for facilitating knowledge transfer and support between Patatrack and industrial partners.

This manuscript has been authored by Fermi Research Alliance, LLC under Contract No. DE-AC02-07CH11359 with the U.S. Department of Energy, Office of Science, Office of High Energy Physics.
\bibliographystyle{JHEP}
\bibliography{ms}

\end{document}